\documentclass[aps,prd,twocolumn,titlepage,superscriptaddress,floatfix]{revtex4}
\usepackage[dvipdfmx]{graphicx}
\usepackage{amsfonts,amsthm,amsmath,amssymb,float,mathtools}
\usepackage[dvipsnames]{xcolor}
\usepackage{subfigure}
\usepackage{color}
\usepackage{appendix}
\usepackage{bbm} %for \mathbbm{1} etc
\usepackage{tensor} % for left subscript
\usepackage{verbatim} % for comments
\usepackage[
	  pagebackref=false,
	  colorlinks=true,
      linkcolor=blue,
      urlcolor=blue,
      filecolor=black,
      citecolor=PineGreen,
      pdfstartview=FitV,
      pdftitle={},
        pdfauthor={},
        pdfsubject={},
        pdfkeywords={},
        pdfpagemode=None,
        bookmarksopen=true
      ]{hyperref}
\usepackage{ascmac}
\usepackage{ulem}
\usepackage{xcolor}
\DeclareRobustCommand{\erase}{\bgroup\markoverwith{\textcolor{red}{\rule[.5ex]{2pt}{0.4pt}}}\ULon}

\graphicspath{ {Pic/} }

\begin{document}

\newtheorem{definition}{Definition}[section]
\newcommand{\be}{\begin{equation}}
\newcommand{\ee}{\end{equation}}
\newcommand{\bea}{\begin{eqnarray}}
\newcommand{\eea}{\end{eqnarray}}
\newcommand{\LE}{\left[}
\newcommand{\R}{\right]}
\newcommand{\nn}{\nonumber}
\newcommand{\Tr}{\text{Tr}}
\newcommand{\N}{\mathcal{N}}
\newcommand{\G}{\Gamma}
\newcommand{\vf}{\varphi}
\newcommand{\LL}{\mathcal{L}}
\newcommand{\Op}{\mathcal{O}}
\newcommand{\HH}{\mathcal{H}}
\newcommand{\arctanh}{\text{arctanh}}
\newcommand{\up}{\uparrow}
\newcommand{\down}{\downarrow}
\newcommand{\ket}[1]{\left| #1 \right>}
\newcommand{\bra}[1]{\left< #1 \right|}
\newcommand{\ketbra}[1]{\left|#1\right>\left<#1\right|}
\newcommand{\rd}{\partial}
\newcommand{\de}{\partial}
\newcommand{\ba}{\begin{eqnarray}}
\newcommand{\ea}{\end{eqnarray}}
\newcommand{\db}{\bar{\partial}}
\newcommand{\we}{\wedge}
\newcommand{\ca}{\mathcal}
\newcommand{\lr}{\leftrightarrow}
\newcommand{\f}{\frac}
\newcommand{\s}{\sqrt}
\newcommand{\vp}{\varphi}
\newcommand{\hvp}{\hat{\varphi}}
\newcommand{\tvp}{\tilde{\varphi}}
\newcommand{\tp}{\tilde{\phi}}
\newcommand{\ti}{\tilde}
\newcommand{\pr}{\propto}
\newcommand{\mb}{\mathbf}
\newcommand{\ddd}{\cdot\cdot\cdot}
\newcommand{\no}{\nonumber \\}
\newcommand{\la}{\langle}
\newcommand{\lb}{\rangle}
\newcommand{\ep}{\epsilon}
 \def\we{\wedge}
 \def\lr{\leftrightarrow}
 \def\f {\frac}
 \def\ti{\tilde}
 \def\ap{\alpha}
 \def\pr{\propto}
 \def\mb{\mathbf}
 \def\ddd{\cdot\cdot\cdot}
 \def\no{\nonumber \\}
 \def\la{\langle}
 \def\lb{\rangle}
 \def\ep{\epsilon}
\newcommand{\mcl}{\mathcal}
 \def\g{\gamma}
\def\Tr{\text{tr}}

\newcommand{\Matsumoto}[1]{{\color{RoyalBlue} \textbf{MM}:~#1}}

%\begin{titlepage}
%\thispagestyle{empty}

%\begin{flushright}

%\end{flushright}
%\bigskip

%\begin{center}
%  \noindent{\large \textbf{Quantum Chaos and Information Scrambling}}\\
%\vspace{2cm}

%\vspace{1cm}
\title{Quantum Chaos, Thermalization, and Non-locality} 
\author{Masataka Matsumoto\footnote{mmatsumoto173@g.chuo-u.ac.jp}}
\affiliation{Department of Physics, Chuo University, Tokyo 112-8551, Japan}
\author{Shuta Nakajima\footnote{nakajima.shuta.qiqb@osaka-u.ac.jp}}
\affiliation{Center for Quantum Information and Quantum Biology, The University of Osaka, Toyonaka, Osaka 560-0043, Japan}
\author{Masahiro Nozaki\footnote{mnozaki@ucas.ac.cn}}
\affiliation{Kavli Institute for Theoretical Sciences, University of Chinese Academy of Sciences,
Beijing 100190, China}
\affiliation{ RIKEN Interdisciplinary Theoretical and Mathematical Sciences (iTHEMS), Wako, Saitama 351-0198, Japan}
\author{Ryosuke Yoshii\footnote{ryoshii@rs.socu.ac.jp}}
\affiliation{Center for Liberal Arts and Sciences, Sanyo-Onoda City University,
1-1-1 Daigaku-Dori, Sanyo-Onoda, Yamaguchi 756-0884, Japan}
\affiliation{International Institute for Sustainability with Knotted Chiral Meta Matter (WPI-SKCM2), Hiroshima University, Higashi-Hiroshima, Hiroshima 739-8526, Japan}

%%%%%%%%%%%%%%%%%%%%%%%%%%%%%
\begin{abstract}
\noindent 
%In this paper, we numerically explore if quantum thermalization occurs during the time evolution induced by the non-local Hamiltonian, the spectra of which exhibit integrability. This non-local Hamiltonian is constructed of two kinds of integrable Hamiltonians. Our findings are as follows: the time dependence of entanglement entropy and mutual information shows that non-locality can evolve the system to the Haar random state; that of logarithmic negativity shows that the non-locality can destroy the quantum correlation. These findings suggest that the quantum thermalization induced by the non-local Hamiltonian does not need the quantum chaoticity of the system.\\

In this paper, we numerically investigate whether quantum thermalization occurs during the time evolution induced by a non-local Hamiltonian
whose spectra exhibit integrability.
This non-local and integrable Hamiltonian is constructed by combining two types of integrable Hamiltonians.
From the time dependence of entanglement entropy and mutual information, we find that non-locality can evolve the system into the typical state.
On the other hand, the time dependence of logarithmic negativity shows that the non-locality can destroy the quantum correlation. These findings suggest that the quantum thermalization induced by the non-local Hamiltonian does not require the quantum chaoticity of the system.
\end{abstract}
%%%%%%%%%%%%%%%%%%%%%%%%

\maketitle
%\vskip 4em
%\end{center}

%\end{titlepage} 
%%%%%%%%%%%%%%%%%%%%%%%%%%%%%%%
%\tableofcontents

%%%%%%%%%%%%%%%%%%%%%%%%%%%%%%%%%%
\section{Introduction \label{sec:intro}}
%%%%%%%%%%%%%%%%%%%%%%%%%%%%%%%%%%

%{\it \color{magenta} Introduction---}
%
%{\it Information scrambling and Quantum chaos-} 
Information scrambling, a non-equilibrium phenomenon where the time evolution locally hides the information about the initial state such as an entanglement structure, is one of the central research subjects in theoretical physics (e.g., see reviews \cite{Lewis2019dynamics,Xu2024scrambling}).
%Information scrambling is a non-equilibrium phenomenon where the time evolution locally hides the information about the initial state, such as the initial entanglement structure. 
After the information scrambling, the thermalization of the subsystems, so-called quantum thermalization, is expected to occur, so that the entanglement structure becomes independent of the initial states.
The complicated quantum systems are expected to cause information scrambling.
A property of the quantum systems, characterizing the ``complexity" of the quantum systems, is called quantum chaoticity. 
Celebrated quantities, measuring this property, are level statistics and spectral form factor (originally discussed in \cite{Berry1977level,Bohigas1984characterization}).
These behaviors are determined by energy spectra of the quantum systems. 
Later the other path to approach the quantum chaos is explored. 
The difficulty of the quantum chaos compared with the classical chaos stems from the absence of the trajectory of the particles which is smeared out by the wave nature. 
The important notion of the classical chaos captured by the Lyapnov exponent, which characterizes the sensitivity of the trajectory on the initial condition, is thus ill-defined. 
In order to avoid this disappearance of the trajectory, for instance, semi-classicalization by continuous measurement is examined in Ref.~\cite{Bhattacharya:1999gx}. 
More recently, a quantity called the out of time ordered correlation (OTOC), is defined to extract the Lyapnov exponent from the quantum system \cite{2016JHEP...08..106M,2015PhRvL.115m1603R}.

Historically, the notion of ``quantum chaos'' has been associated with the level statistics of the energy spectrum; the quantum integrability of systems is known to be reflected in the distribution of level spacing.
The random matrix theory certifies that the level spacing statistics follow a Wigner-Dyson distribution in chaotic systems, whereas a Poisson distribution in integrable systems. 
Nevertheless, the spectral definition of quantum chaos turns out to be subtle because, for example, it becomes non-trivial and hard to treat in the thermodynamic limit.

Numerous researchers have been exploring a significant relationship between information scrambling (quantum thermalization) and quantum chaos \cite{Berry1977level,2006NatPh...2..754P,2006PhRvL..96e0403G,2017AnP...52900301G,2015PhRvL.115j0402G,2010PhRvE..81a1109G,2016JSP...163..937T,2006cond.mat..2625S,Bohigas1984characterization,1977JPhA...10.2083B,1986PhRvA..34..591F,2016AdPhy..65..239D,2017PhRvE..96a2157M,1989PhyS...40..335B,2004PhRvL..93a4103M,2005PhRvE..72d6207M,2018PhRvX...8b1062K,2018PhRvL.121z4101B,2020PhRvE.101e2201B,2017PhRvX...7c1016N,2018PhRvX...8b1013V,2018PhRvL.121f0601C,2019PhRvL.123u0603F,2019PhRvX...9b1033B,2018PhRvX...8d1019C,1993PhRvL..71.1291P,1994PhRvE..50..888S,1996PhRvL..77....1S,1999JPhA...32.1163S,2011PhRvL.106e0405B,2017JHEP...05..118C,2018JHEP...07..124G}. 
For example, they have been studying if the emergence of information scrambling needs the quantum chaoticity of the systems. 
%\footnote{\Matsumoto{Our question: Is the information scrambling a sufficient condition for quantum chaos? Can we control the scrambling, describing the local dynamics, without changing the global energy spectra of a system?}}. 

%{\bf Counter-example} 
In this paper, we explore if quantum thermalization (information scrambling) occurs during the time of evolution induced by a non-local Hamiltonian that does not exhibit quantum chaos.

 The non-local Hamiltonians considered in this paper are 
\be
H_{\text{NL}}(\tau)=u^{-1}(\tau) H u(\tau),
\label{Hdeform}
\ee 
where we assume that $\tau$ is a complex value and the transformation, $u(\tau)$, satisfies that $u(\tau)u^{-1}(\tau)=u^{-1}(\tau)u(\tau)={\bf 1}$, where $\bf 1$ is an identity operator. 
Furthermore, we assume that $H$ is a Hermitian operator, $H^{\dagger}=H$, where $\dagger$ is the Hermitian conjugation. If $[H,u(\tau)]=0$, $H_{\text{NL}}(\tau)$ reduces to the original Hamiltonian, $H$.
If the transformation is unitary, $u^{-1}(\tau)=u^{\dagger}(\tau)$, the non-local Hamiltonians are also Hermitian.
Otherwise, $H_{\text{NL}}(\tau)$ is a non-Hermitian operator.
Here, $\tau$ is the parameter controlling the non-locality of the Hamiltonians: For small $\tau$, this Hamiltonian is approximately given by $H$, while for large $\tau$, the non-local terms created by $u(\tau)$ can significantly contribute to $H_{\text{NL}}(\tau)$.
In Section {\ref{sec:GP-NL-H}}, we will report on the fundamental properties of these non-local Hamiltonians.

{\it  Summary---}
As a simple example, we considered the non-local Hamiltonian, $H_{\text{NL}}(\tau)$, the spectra of which exhibit the integrability 
%(See (\ref{eq:Hamiltonians-considered}) for its details ) 
(See Sec.~\ref{sec:GP-NL-H} for its details )
and the chaotic Ising model $H_{\text{CI}}$ as a reference system. 
Here, $u(\tau)$ is the unitary operator, $u(\tau)=e^{-i \tau H_1}$, where $H_1$ is the Hermitian operator.
We explored if the non-locality of $H_{\text{NL}}(\tau)$ induces the quantum thermalization. 
More precisely, by studying the time dependence of entanglement entropy and mutual information, we explored how closely $H_{\text{NL}}(\tau)$ drives the subsystems to the typical state. Here the typical state means that those entanglement measures coincide with those for the Haar random average \cite{1993PhRvL..71.1291P,1996PhRvL..77....1S}.
Then, we can see from $\tau$ dependence of their late-time behaviors that the larger $\tau$ becomes, the closer the subsystems approaches the typical state.
%{\color{red}This suggests that the non-locality of $H_{\text{NL}}(\tau)$ prevents the system from obtaining the quantum properties\sout{, such as quantum recursion and quantum non-local correlation.} }%\textcolor{blue}{This behavior of mutual information suggests that non-locality can destroy the quantum correlation.} 
This suggests that the non-locality of $H_{\text{NL}}(\tau)$ time-evolves the subsystems to the typical state.
Furthermore, to check it, we explored the time dependence of the logarithmic negativity, and then this time dependence shows that the non-locality of $H_{\text{NL}}(\tau)$ can prevent the system from obtaining the quantum correlation.

{\it Organization of this paper---}
In Section \ref{sec:GP-NL-H}, we will explain the general properties of the non-local Hamiltonians considered in this paper.
Then, in Section \ref{sec:Ising-system}, we will explain the setup considered in this paper, and report our findings.
Finally, in Section \ref{sec:discussions-and-future}, we will close this paper with comments on our findings and future directions.

\begin{comment}
%%%%%%%%%%%%%%%%%%%%%%%%%%%%%%%%%%%
\subsection*{Organization of this paper}
%%%%%%%%%%%%%%%%%%%%%%%%%%%%%%%%%%%
In Section \ref{sec:properties-of-DH}, we will describe the fundamental property of the generic deformed Hamiltonian.
In Section \ref{sec:Ising-system}, we will explore the dynamics induced by the deformed simple Ising Hamiltonian.
In Section \ref{sec:discussions-and-future}, we will discuss our findings in this paper.
\end{comment}

%{\it \color{magenta} General properties of the non-local Hamiltonian ---}
%%%%%%%%%%%%%%%%%%%%%%%%%%%%%%%%%%
\section{General properties of non-local Hamiltonians \label{sec:GP-NL-H}}
%%%%%%%%%%%%%%%%%%%%%%%%%%%%%%%%%%
Before moving to the analysis of the specific system considered in this paper, we will report the general properties of the non-local Hamiltonians.
%%%%%%%%%%%%%%%%%%%%%%%%%%%%%%%%%%
\subsubsection*{Energy spectra of the non-local Hamiltonians}
%%%%%%%%%%%%%%%%%%%%%%%%%%%%%%%%%%
The eigenvalues of $H_{\text{NL}}(\tau)$ are the same as those of $H$,
\be \label{eq:global-property-one}
\begin{split}
\text{det}\left(H_{\text{NL}}(\tau)-\lambda\bf{1}\right)=\text{det}\left(H-\lambda \bf{1}\right)=0,
\end{split}
\ee
where ${\bf 1}$ is an identity operator.
This implies that $u(\tau)$ does not change the global quantities, depending only on the eigenvalues of the Hamiltonian, such as the spectral form factor, level spacing, resolvent, and so on. 
In other words, the global property of the non-local Hamiltonian is the same as that of the original one. 
In addition, suppose that the system is in the thermal state  governed by $H_{\text{NL}}(\tau)$, $\rho=e^{-\beta H_{\text{NL}}(\tau)}/\Tr e^{-\beta H_{\text{NL}}(\tau)}$, where the inverse temperature is $\beta$,  the thermodynamic properties, such as thermal entropy, of this system are the same as those of the thermal system in $\rho_0=e^{-\beta H}/\Tr e^{-\beta H}$.
Let $\ket{E_a}$ denote the eigenstates of $H$.
Here, $a$ labels the eigenvalues of $H$.
Then, the bra- and ket- eigenstates of $H_{\text{NL}}(\tau)$ are given by $\bra{E_a} u(\tau)$ and $u(\tau)^{-1} \ket{E_{a}}$, respectively.
 
%%%%%%%%%%%%%%%%%%%%%%%%%%%%%%%%%%
\subsubsection*{Non-locality}
%%%%%%%%%%%%%%%%%%%%%%%%%%%%%%%%%%  
Here, we discuss the non-locality, induced by $u(\tau)=e^{-i \tau H_1}$, of $H_{\text{NL}}(\tau)$.
Let us define the local Hamiltonians as the ones containing only the nearest neighbor %\sout{- and self-interactions.} 
interaction and the onsite term. 
For simplicity, we assume that $u(\tau)$ is a unitary operator, $u(\tau)=e^{-i\tau H_1}$, where $H_1$ is a local Hermitian.
Even if we start from the local Hamiltonian, $H$, $u(\tau)$ can make it a non-local one.
%\textcolor{blue}{\sout{Furthermore, we assume that $H$ and $H_1$ are constructed of only local terms, $H=\sum_{i}h^{i}$ and $H_1=\sum_{i}h^{i}_1$, where $h^i$ and $h^i_1$ are Hermitian, and they act on $i$-th and $i+1$-th sites.}}
The Baker–Campbell–Hausdorff formula reduces $H_{\text{NL}}(\tau)$ to the nests of the commutators of $H$ and $H_1$,
\be
H_{\text{NL}}(\tau)=H+(i\tau)\left[H_{1},H\right]+\f{(i\tau)^2}{2!}\left[ H_{1},\left[H_{1},H\right]\right]+\cdots.
\ee
Here, the commutators can create the non-local terms, acting not only on $i$-th and nearest neighbor-sites, but also on %\textcolor{red}{$i+(j>2)$-th site}$\rightarrow$ 
further far sites from $i$-th one.
The larger the number of commutators is, the more non-local terms can be induced. %the non-local Hamiltonian possesses. 
For the large $\tau$, the terms with a large number of commutators dominantly contribute to $H_{\text{NL}}(\tau)$%\sout{.} 
%\textcolor{red}{\sout{In other words, for the large $\tau$, the non-local terms can significantly contribute to the dynamics.}} 
, and thus the non-local terms can significantly contribute to the dynamics.

%%%%%%%%%%%%%%%%%%%%%%%%%%%%%%%%%%
\subsubsection*{Local property of the non-local Hamiltonians}
%%%%%%%%%%%%%%%%%%%%%%%%%%%%%%%%%%
The local quantities, such as two-point function, entanglement entropy, and so on, depend on the transformation $u(\tau)$. 
%\textcolor{magenta}{The reason is as follows; the expectation value of the local observable $\mathcal{O}_i$ is calculated by the reduced density matrix $\Tr{\rho \mathcal{O}_i}=\Tr_i [(\Tr_{\bar i}\rho) \mathcal{O}_i]$, where the partial trace is not commutable with the transformation by $u(\tau)$. This non-commutability also results in the $u(\tau)$ dependence of the entanglement entropy.} 
As a simple example, let us consider the $n$-point function of the thermal state determined by $H_{\text{NL}}(\tau)$.
Define this $n$-point function as $\langle \prod_{i}\mathcal{O}_i\rangle_{H_{\text{NL}}}:=\Tr\left( e^{-\beta H_{\text{NL}}}\mathcal{O}_1\cdots \mathcal{O}_n\right)/\Tr e^{-\beta H_{\text{NL}}}$.
We assume $[u(\tau),\mathcal{O}_i] \neq 0$ and define the local operators in Heisenberg's picture as $\mathcal{O}_{i,u}(\tau):=u(\tau)\mathcal{O}_{i}u^{-1}(\tau)$.
Then, the $n$-point function reduces to
\be \label{eq:local-property}
\begin{split}
    \left\langle \prod_{i}\mathcal{O}_i\right\rangle_{H_{\text{NL}}}
    %&\erase{=\f{\Tr\left(e^{-\beta H_{2}}  \mathcal{O}_{1,H_1}(\tau)\cdots \mathcal{O}_{n,H_1}(\tau) \right)}{\Tr e^{-\beta H_{2}}}}\\
    &= \left\langle \prod_{i}\mathcal{O}_{i,u}(\tau)\right\rangle_{H}.
\end{split}
\ee
%where $u_1=\exp(iH_1 \tau)$. 
Generically, $\left\langle \prod_{i}\mathcal{O}_{i,u}(\tau)\right\rangle_{H}\neq \left\langle \prod_{i}\mathcal{O}_{i,u}(\tau=0)\right\rangle_{H}$.
%Thus, the local quantities, including $n$-point functions, can distinguish between $e^{-\beta H}$ and $e^{-\beta H_{\text{NL}}}$.
%\textcolor{blue}{One interpretation of (\ref{eq:local-property}) is that the local behavior of the system depends on the non-locality of $H_{\text{NL}}(\tau)$.} $\Rightarrow$ 
One interpretation of (\ref{eq:local-property}) is that $u(\tau)$, that induces the non-locality of $H_{\text{NL}}(\tau)$, delocalizes the local operators, $\mathcal{O}_i$.

%In the following, we mainly consider the case with $[H_1,H_2]\neq 0$. 
%We also assume that $H_i$ are Hermitian operators, $H^{\dagger}_i=H_i$. 
%For the specific model, we considered the spin systems in this paper. 
%These Hamiltonians have the following properties:
%(1) They are Hermitian irrespective of $\tau$. 
%(2) When $\tau$ becomes larger, more non-local terms can emerge and contribute to the dynamics even if $H_i$ are local. 
%(3) The energy spectrum of $H_{\text{eff}}$ is equivalent to that of $H_2$,
%\be
%\text{det}\left(H_{\text{eff}}(\tau)-\lambda\bf{1}\right)=\text{det}\left(e^{iH_1\tau} H_{2} e^{-iH_1\tau}-\lambda e^{iH_1\tau}  e^{-iH_1\tau}\right)=\text{det}(e^{iH_1\tau}e^{-iH_1\tau})\text{det}\left(H_{2}-\lambda \bf{1}\right)=0.
%\ee

%%%%%%%%%%%%%%%%%%%%%%%%%%%%%%%%%%
%\subsubsection*{\sout{Symmetry}}
%%%%%%%%%%%%%%%%%%%%%%%%%%%%%%%%%%
%\erase{Now we conclude this section by commenting on the symmetry of the deformed Hamiltonian.
%We assume that the system has some conserved charge $Q$, which commutes with $H$.
%In this case, we can define the conserved charge, $Q_{\text{def}}$, for $H_{\text{NL}}$ as $Q_{\text{def}}=u(\tau)Q u^{-1}(\tau)$.  }

%%%%%%%%%%%%%%%%%%%%%%%%%%%%%%%%%%
\section{Setup and our findings \label{sec:Ising-system}}
%%%%%%%%%%%%%%%%%%%%%%%%%%%%%%%%%%
%{\it \color{magenta} Setup---}
Here, we will present the setup considered in this paper and our findings.         
%\textcolor{red}{For demonstration, we use two transverse Ising models} $\rightarrow$ 
The non-local Hamiltonian considered in this paper is constructed of two kinds of transverse Ising Hamiltonians with the open boundary condition,
\begin{align} \label{eq:Hamiltonians-considered}
%&H=\sum_{i=1}^{N}\left(-Js^{(i)}_zs^{(i+1)}_z+h_xs^{(i)}_x\right),\\
%&H_1=\sum_{i=1}^{N}\left(-Js^{(i)}_zs^{(i+1)}_z+h_ys^{(i)}_y\right),
&H=H_{\rm int}+\sum_{i=1}^{N}h_xs^{(i)}_x,\ 
H_1=H_{\rm int}+\sum_{i=1}^{N}h_ys^{(i)}_y,
\end{align}
where, $s^{(i)}_{\gamma=x,y,z}$ and $N$ denote the Pauli matrices acting on $i$-th site, and the total number of the system sites, respectively. 
Here, $H_{\rm int}=-J\sum_{i=1}^{N-1}s^{(i)}_zs^{(i+1)}_z$ is the interaction part. 
In the following, we set $\hbar=1$, $J=1$, and $h_x=h_y=1$ for simplicity. 
It is known that those transverse Ising model is integrable \cite{PFEUTY197079}. 
The scale of the time is set to be $\hbar/J$. 
Since the dimension of the Hilbert space becomes $2^N$ for an $N$-site system, we frequently use $2$ as the base of the logarithm and omit that in the whole manuscript, namely $\log \alpha$ means $\log_2 \alpha$. 
In the later section, we also use the chaotic Ising (CI) model given by 
\begin{align} \label{eq:chaosIsing}
&H_{\rm CI}=H_{\rm int}+\sum_{i=1}^{N}\left(\tilde h_xs^{(i)}_x+\tilde h_zs^{(i)}_z\right). 
\end{align}
In the following, the values of $\tilde h_x$ and $\tilde h_z$ are set to be $1.05$ and $-0.5$, respectively. 
The system considered in this paper is in the time-evolved pure state, 
\be
\ket{\Psi(t,\tau)}=e^{-itH_{\text{NL}}(\tau)}\ket{\Psi}
\ee
where we assume that all spins are up in the initial normalized state, $\ket{\Psi}$.
The parameter, $t$, is the time associated with $H_{\text{NL}}(\tau)$.%$H_{\text{non}(\tau)}$

For later convenience, we labels subsystems in two ways. 
In the case of the bipartite partition, we refer to the subsystem as $A$ and rest of the system as $\overline A$ (inset in Fig.\ref{EE4site}). 
On the other hand, when we consider the tripartite partition, we label as the left ($A$), the middle ($C$), and the right ($B$) subsystem (inset in Fig. \ref{MI1site}). 
%Note that we only consider the tripartite partition for the open boundary condition and thus the left and the right subsystem is disconnected. 

In this setup, we explore the $\tau$ dependence of entanglement dynamics induced by $H_{\text{NL}}(\tau)$.

%%%%%%%%%%%%%%%%%%%%%%%%%%%%%%%%%%
\subsection*{Our findings}
%%%%%%%%%%%%%%%%%%%%%%%%%%%%%%%%%%
We will report the $\tau$-dependence of the time evolution of the entanglement entropy, mutual information, and logarithmic negativity for $N=10$ (for results of different sizes, see Appendix A). 
%%%%%%%%%%%%%%%%%%%%%%%%%%%%%%%%%%
\subsubsection*{Entanglement entropy}
%%%%%%%%%%%%%%%%%%%%%%%%%%%%%%%%%%
We will begin by explaining the definition of the entanglement entropy, the quantity that measures the bipartite entanglement.
Suppose that the system considered is in the state described by the density matrix $\rho$. %$\ket{\psi}$, and then define a density operator as $\rho=\ket{\psi}\bra{\psi}$.
We spatially divide the Hilbert space into the subsystems $A$ and  $\overline{A}$.
Then, define the reduced density matrix associated with $A$ as 
\be
\rho_{A}= \Tr_{\overline{A}} \rho.
\ee
By using the reduced density matrix, entanglement entropy is defined as follows, 
\be
S_{A}=-\Tr_{A}\rho_{A}\log{\rho_A}.
\ee

\begin{figure}[tbp]
    \includegraphics[width=0.9\columnwidth]{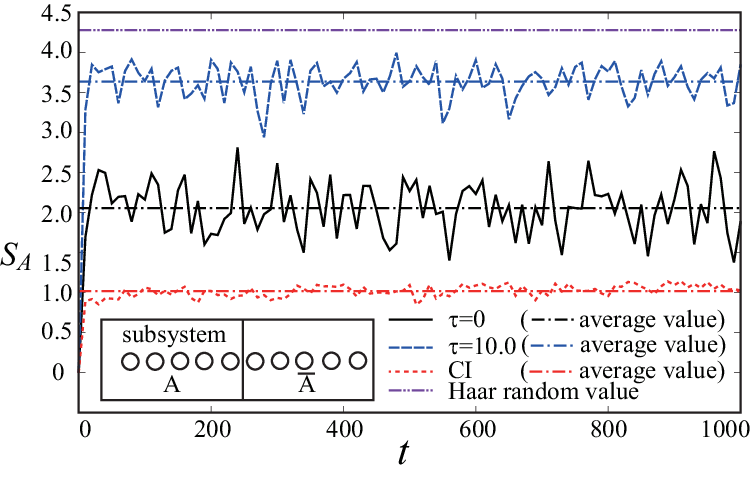}
    \caption{Time dependence of the entanglement entropy for $\tau=0$ (solid line), $\tau=10$ (dashed line), and for the CI model (dotted line). 
    The choice of the subsystem in this calculation is shown in the inset. 
    The average values over the range from $t=0$ to $t=1000$ are also shown in the figure (dash-dotted lines). 
    The average values are $\sim$  2.056 ($\tau=0$), $\sim$ 3.636 ($\tau=10$), and $\sim$ 1.019 (CI model). 
    The variances are $\sim 0.1006$, $\sim 0.04197$, and $\sim 0.004928$ for $\tau=0$, $\tau=10$, and the CI model, respectively. 
    We also show the value calculated by the case of the random pure state $\sim$ 4.279 (dash-dot-dot line). 
    We note that though the lines are shown to make it visually clear, calculation is done for 100 points ($t=0,10,20,\cdots$) in each case. 
    }
    \label{EE4site}
\end{figure}

Subsequently, we will show the $\tau$ dependence of the entanglement entropy.
In Fig.\ \ref{EE4site}, we plot the time dependence, the average value, and variance of  the entanglement entropy for $\tau=0$, $\tau=10$, and the CI model (to avoid the effect from the initial relaxation, we take the time average removing the data for initial time throughout this study (for technical detail, see Appendix B)).  %\footnote{Namely, we calculate the mean value of $\hat A(t)$ by ${\overline A}=\frac{1}{T}\sum_{n=1}^{T/\Delta t}\langle \hat A(n\Delta t)\rangle $ and variance to be $\frac{1}{T}\sum_{n=1}^{T/\Delta t}\langle ({\overline A}-\hat A(n\Delta t))^2\rangle$. See the supplemental material for more details.}. 
Here, we choose the size of the subsystem to be 5 (for schematic picture, see the inset of Fig.\ \ref{EE4site}).
The time dependence of entanglement entropy exhibits the oscillatory behavior in time.
For $H_{\text{NL}}(\tau=0)$, the oscillation with time may be due to its integrability. The average value for $\tau=0$ is larger than that for the CI model. However, since the variance for $\tau =0$ is larger than the others, the mutual information may not saturate approximately the average value.
%Similarly, we choose two adjacent sites, three adjacent sites, and four adjacent sites at edge are chosen to be the subsystem for top-right, bottom-left, and bottom-right panels, respectively. 
It is shown that the recursive behaviour of the entanglement entropy is suppressed for $\tau=10$, compared with $\tau=0$.  
From Fig.\ \ref{EE4site}, we can see that in the case of $\tau=10$, the average value of entanglement entropy almost coincides with the Haar random value and the fluctuation becomes much smaller than the case of $\tau=0$. 
Here, the Haar random value is defined as the average entanglement entropy over the random pure states $S_A^{\rm HR}$ \cite{1993PhRvL..71.1291P,1996PhRvL..77....1S} given by 
\begin{equation}
S_A^{\rm HR}=\sum_{k=d_A}^{d_{A}d_{\bar A}}\frac{1}{k}-\frac{d_A-1}{2d_{\bar A}}, 
\label{HaarRandom}
\end{equation}
where $d_A$ and $d_{\bar A}$ are the dimensions of the Hilbert spaces for subsystems $A$ and $\bar A$, respectively. 
 %(throughout this study, we take the time average removing the data for initial time to avoid the effect from the initial relaxation). 
These behaviours can be interpreted as the emergence of the quantum thermalization (information scrambling) due to the non-locality. 
%The saturation value becomes more further apart from the maximum value when the size of the subsystem becomes larger. This behaviour is known as the Page-curve in the case of the finite system, and thus it does not imply the scrambling is not achieved for larger subsystem sizes. 
We note that this 
%scrambling 
quantum thermalization is not induced by the quantum chaos in the sense that the energy level statistics do not show the Wigner-Dyson level statistics.

It would be noted that the spin models with the second nearest neighbor coupling and chaotic SYK model are known to exhibit the thermalization and the scrambling \cite{Iyoda:2017pxe,Yoshii:2020fog,2024PhRvR...6b2021N}.
From the view point of our present work, the key property of those models for the thermalization and the scrambling would be the non-locality.

%%%%%%%%%%%%%%%%%%%%%%%%%%%%%%%%%%
\subsubsection*{Mutual information}
%%%%%%%%%%%%%%%%%%%%%%%%%%%%%%%%%%
Next, we will explain the definition of the mutual information, the quantity measuring the non-local correlation.
This quantity contains both classical and quantum correlations \cite{PhysRevA.72.032317}. 
To define it, spatially divide the Hilbert space into $A$, $B$, and  $\overline{A\cup B}\equiv C$. 
Then, define the mutual information between $A$ and $B$ as 
\be
I_{A,B}=S_A+S_B-S_{A\cup B}.
\ee
%\textcolor{red}{In the previous subsection, we explored the $\tau$ dependence of the entanglement entropy. Though the entanglement entropy characterizes the entanglement between the subsystem and the rest part of the system, it is known that the entanglement entropy consists of the classical correlation and the quantum correlation. In this part, we consider the mutual information and the negativity between a site at the edge and another site at the other edge. Both quantities are considered to characterize the quantum correlation between two subsystems}\footnote{\textcolor{red}{This sentence looks wrong because the mutual information measures the quantum and classical correlation. }}. 
%\textcolor{red}{It would be expected if the scrambling takes place, the quantum correlation between subsystems apart from each other vanishes.}$\rightarrow$
In the previous section, we explore the time dependence of the single interval entanglement entropy for $\tau=0$, $\tau=10$, and the CI model. 
Then, we found that during the time evolution induced by $H_{\text{NL}}(\tau)$ with $\tau=10$, the subsystems %\sout{approximately} 
evolve in time to the typical state. %\sout{Haar random state} typical state. 
Here, we will explore if the late-time value of the mutual information results in that for the typical state.

In Fig.\ \ref{MI1site}, we report the time evolution, the average value, and the variance of mutual information for $\tau=0$, $\tau=10$, and the CI model. 
The mutual information also exhibits the oscillatory behavior with time. The time dependence and the variance of mutual information in Fig.\ \ref{MI1site} show that the fluctuation in time for $\tau=0$ is much larger than those for $\tau=10$ and the CI model.
Hence, the large-time value of mutual information for $\tau=0$ may not be approximated by the average value.
The average value of the mutual information for $\tau=10$ is closer to the Haar random value than that for the CI mode, which suggests that the non-local correlation for $\tau=10$ may result in that for the typical state. 
%The left panel shows the short time dynamics. 
%Since the Hamiltonian becomes more non-local when $\tau$ becomes larger, the time delay for the initial growth, which is clearly seen in the case of $\tau$, vanishes. 
%Moreover, the magnitude of the mutual information also becomes small. 
%Especially, the latter implies that the correlation between two subsystems far apart is scrambled. 
%We also show the long time behaviour of the mutual information in the right panel of Fig.\ \ref{MI1site}. 
%It is shown that the average value indeed decreases as $\tau$ increases and the fluctuation is suppressed for larger $\tau$. 
%%%%%%%%%%%%%%%%%%%%%%%%%%%%%%%%%%
\subsubsection*{Logarithmic negativity}
%%%%%%%%%%%%%%%%%%%%%%%%%%%%%%%%%%
%In the previous section, we explored how the non-local correlation measured by $I_{A,B}$ depends on $\tau$.
%Consequently, we saw that strong non-locality destroys the non-local correlation.

In this section, we will closely look at $\tau$ dependence of quantum correlation by using the logarithmic negativity. 
Let us define the logarithmic negativity, proposed as the quantity that measures the quantum correlation \cite{vidal2002computable,plenio2005logarithmic,PhysRevLett.77.1413,HORODECKI19961}. 
This property depends solely on the separability of the quantum system. 
Consider the reduced density matrix associated with $A\cup B$, namely $\rho_{A\cup B}\equiv \rm{Tr}_C\rho$, and then define the partial transpose, acting on the region $B$, as
\be
\left(\rho^{T_B}_{A\cup B~}\right)_{ij;ab} =\left(\rho_{A\cup B}\right)_{ij;ba}
\ee
where $i$ and $j$ denote the ket- and bra-states associated with $A$, while $a$ and $b$ denote the ket- and bra-states associated with $B$.
Subsequently, define the logarithmic negativity as 
\be
\mathcal{E}_{A,B}=\log{\Tr\left|\rho^{T_B}_{A\cup B}\right|},
\ee
where $\Tr\left|\rho^{T_B}_{A\cup B}\right|$ stands for 
$\sum_{l}\left|\lambda^{T_B}_{l}\right|$, where $\lambda^{T_B}_{l}$ are eigenvalues of $\rho^{T_B}_{A\cup B}$.

\begin{figure}[tbp]
    \centering
    \includegraphics[width=1\columnwidth]{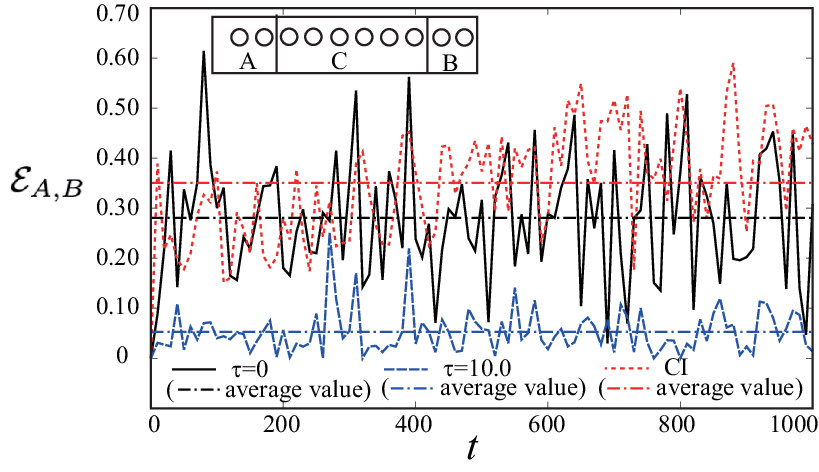}
    \caption{The logarithmic negativities as functions of the time for $\tau=0$ (solid line), $\tau=10$ (dashed line), and for the CI model (dotted line). 
    The partition for this calculation is shown in the inset. 
    The average value is shown in the right figure by the horizontal dash-dotted lines. 
    The average values are, respectively, $\sim$ 0.2807 ($\tau=0$), $\sim$ 0.05270 ($\tau=10$), and $\sim$ 0.3503 (CI model).
    The variances are $\sim 0.01421$, $\sim 0.001885$, and $\sim 0.01092$ for $\tau=0$, $\tau=10$, and the CI model, respectively.
    }
    \label{LogNeg1site}
\end{figure}

%\textcolor{red}{We also calculate the logarithmic negativity since it is known that this quantity becomes nonzero only if the quantum correlation is present.}
%\sout{Then, we will report the $\tau$ dependence of the logarithmic negativity.}
In Fig.\ \ref{LogNeg1site}, we show the time dependence, the average value, and the variance of the logarithmic negativities for $\tau=0$, $\tau=10$, and the CI model as functions of the time.
The time dependence of the logarithmic negativities for them exhibits the oscillatory behavior with time. Since the variance of the logarithmic negativity for $\tau =0$ is larger than the others, the late-time value may not be approximated by the average one. Since the average value for $\tau =10$ is smaller than the CI model, we can see that the quantum correlation is highly suppressed for $\tau=10$ which has the strong non-locality. 
%It clearly shows that the quantum correlation is highly suppressed for $\tau=10$ which has the strong non-locality. 
This result implies that the scrambling prevents the growth of the quantum correlation between two distant subsystems.

%It is consistent with the physical interpretation drawn from the results of the mutual information. 

\begin{figure}[tbp]
    \centering
    \includegraphics[width=0.9\columnwidth]{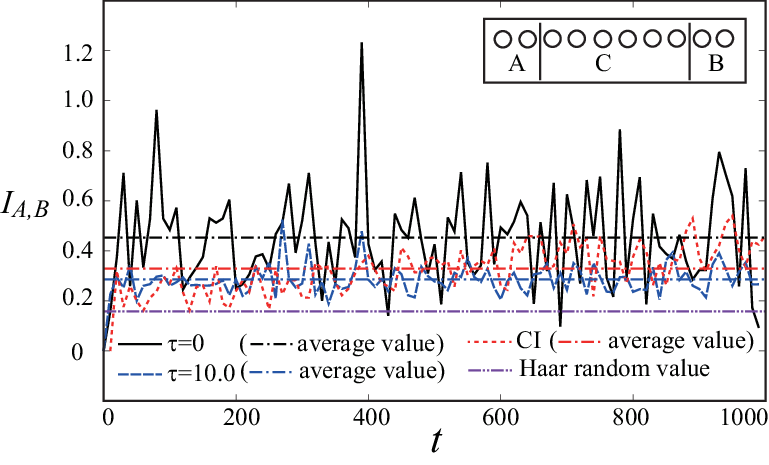}
    \caption{Time dependence of the mutual information. 
    The partition for this calculation is shown in the inset. 
    The solid line, the broken line, and dotted line corresponds to $\tau=0$, $10$, and the CI model, respectively. 
    The average values are shown in the right figure by the horizontal dash-dotted line. 
    The value calculated by the case of the random pure state is also shown (dash-dot-dot line). 
    The average values are, respectively, $\sim$ 0.4525 ($\tau=0$), $\sim$ 0.2857 ($\tau=10$), and $\sim$ 0.3287 (CI model). 
    The variances are $\sim 0.003681$, $\sim 0.003139$, and $\sim 0.002907$ for $\tau=0$, $\tau=10$, and the CI model, respectively.
    We also show the value calculated by the case of the random pure state ($\sim$ 0.1585). }
    \label{MI1site}
\end{figure}

\if[0]
%%%%%%%%%%%%%%%%%%%%%%%%%%%%%%%%%%
\section{\sout{Results}}
%%%%%%%%%%%%%%%%%%%%%%%%%%%%%%%%%%
\erase{In this section, we will show the results of the aforementioned quantities computed from the non-local Hamiltonian.}
%{\it \color{magenta} Results---}
\subsection{\sout{Entanglement entropy}}
\erase{In this part we show the $\tau$ dependence of the entanglement entropy. 
Since $H_{\text{NL}}(\tau=0)$ is integrable, recursion appears. 
Moreover, integrable system does not achieve the scrambling and thus the entanglement entropy does not saturates the maximum value. 
In Fig.\ \ref{EE1site}, we plot the $\tau$ dependence of the entanglement entropy. 
For the top-left panel, we choose a site at the edge to be the subsystem. 
Similarly, we choose two adjacent sites, three adjacent sites, and four adjacent sites at edge are chosen to be the subsystem for top-right, bottom-left, and bottom-right panels, respectively. 
It is shown that the recursive behaviour of the entanglement entropy is suppressed by introducing $u(\tau)$. 
In the case of $\tau=10$, the entanglement entropy almost saturates the maximum value and the fluctuation becomes much smaller than the case of $\tau=0$ or $\tau=1$. 
These behaviour can be interpreted as the emergence of the scrambling due to the deformation. 
The saturation value becomes more further apart from the maximum value when the size of the system becomes larger. This behaviour is known as the Page-curve in the case of the finite system, and thus it does not imply the scrambling is not achieved for larger subsystem sizes. 
We note that this scrambling is not induced by the quantum chaos in the sense that the energy level statistics does not show the Wigner-Dyson level statistics. }

\subsection{\sout{Mutual information}}
\erase{In the previous subsection, we examine the $\tau$ dependence of the entanglement entropy. 
Though the entanglement entropy characterizes the entanglement between the subsystem and the rest part of the system, it is known that the entanglement entropy consists of the classical correlation and the quantum correlation. 
In this part, we consider the mutual information and the negativity between a site at the edge and another site at the other edge. 
Both quantities are considered to characterize the quantum correlation between two subsystems. 
It would be expected if the scrambling takes place, the quantum correlation between subsystems apart from each other vanishes. 

In Fig.\ \ref{MI1site}, we plot the mutual information for various $\tau$s. 
The left panel shows the short time dynamics. 
Since the Hamiltonian becomes more non-local when $\tau$ becomes larger, the time delay for the initial growth, which is clearly seen in the case of $\tau$, vanishes. 
Moreover, the magnitude of the mutual information also becomes small. 
Especially, the latter implies that the correlation between two subsystems far apart is scrambled. 
We also show the long time behaviour of the mutual information in the right panel of Fig.\ \ref{MI1site}. 
It is shown that the average value indeed decreases as $\tau$ increases and the fluctuation is suppressed for larger $\tau$. }

\subsection{\sout{Logarithmic negativity}}
\erase{We also calculate the logarithmic negativity since it is known that this quantity becomes nonzero only if the quantum correlation is present. 
In Fig.\ \ref{LogNeg1site}, we plot the logarithmic negativities for various $\tau$ as functions of the time. 
It clearly shows that the quantum correlation is no longer generated when the finite $\tau$ induces the non-locality. 
This result again shows that the scrambling invents the growth of the quantum correlation between two subsystem.  
It is consistent with the physical interpretation drawn from the results of the mutual information. }

\fi
%\clearpage

%%%%%%%%%%%%%%%%%%%%%%%%%%%%%%%%%%
\section{Discussions and future directions \label{sec:discussions-and-future}}
%%%%%%%%%%%%%%%%%%%%%%%%%%%%%%%%%%
%{\it \color{magenta} Discussion---}
We will close this paper with comments on the future problems and the experimental realizations. 
{\it Symmetry of the non-local Hamiltonians}: 
There can exist the symmetry of only $H_{\text{NL}}(\tau)$. This means that $[H,Q]\neq0$, but $[H_{\text{NL}}(\tau),Q]=0$, where $Q$ is the charge associated with the symmetry of $H_{\text{NL}}(\tau)$. 
%\textcolor{magenta}{In the present system, one can easily construct such conserved quantity for $H_{\text{NL}}(\tau)$ operators via $u^{-1}(\tau)\hat O u(\tau)$, where $\hat O$ is the conserved quantities for $H$, namely $[H_{\text{NL}}, u^{-1}(\tau)\hat O u(\tau)]=u^{-1}[H,Q]u(\tau)=0$ whereas $[H,u^{-1}(\tau)\hat O u(\tau)]\neq 0$ .} 
Contrary to this, some symmetries of $H$ cannot be those of $H_{\text{NL}}(\tau)$. 
It would be interesting to explore the dynamics associated with such symmetries.
{\it Quantum chaotic system with no occurrence of the quantum thermalization}: 
In this paper, we found the integrable system exhibiting quantum thermalization.
As a next step, it would be interesting to find the quantum chaotic system without the occurrence of quantum thermalization.
Such systems will suggest that quantum chaoticity is not strongly related to the emergence of quantum thermalization, at least for the non-local Hamiltonian.
{\it Non-Hermitian systems}:
As explained earlier, $H_{\text{NL}}(\tau)$ can be non-Hermitian. Let us consider the thermal state determined by $H_{\text{NL}}(\tau)$. The thermodynamic quantities of this system are independent of $u(\tau)$, while the local quantities, including entanglement entropy, can. 
Such systems may open a new research field of non-Hermitian systems.

%%%%%%%%%%%%%%%%%%%%%%%%%%%%%%%%%%
\subsubsection*{Experimental feasibility}
%%%%%%%%%%%%%%%%%%%%%%%%%%%%%%%%%%
The time evolution of the state $|\psi\rangle $ by the non-local Hamiltonian is given by $e^{-iH_{\text{NL}}(\tau)t}|\psi\rangle =e^{iH_1 \tau}e^{-iH t}e^{-iH_1 \tau}|\psi\rangle$. 
This can be realized by subsequently evolving the system by $H_1$, $H$, and $-H_1$ during time duration $\tau$, $t$, and $\tau$, respectively. 
%%%%%%%%%%%
The Ising interaction part $H_{\rm int}$ whose sign of coupling $J$ is reversible ($J \to -J$) can be implemented in analog quantum simulators, such as those using trapped ions \cite{garttner2017measuring} or superconducting qubits \cite{SCqubit}. In such qubit systems, the transverse field term $h_x s_x^{(i)}$ can be implemented using Rabi coupling between two spin states. %proportional to Pauli $s_x$. 
The transverse field term for the $y$- or $-y$- direction can be effectively implemented by 
inserting appropriate rapid rotation around the $z$-axis, which does not affect the $s_z^{(i)}$ term, before and after $h_x s_x^{(i)}$ dynamics. These kinds of analog quantum simulators based on qubit systems are also suitable for the correlation measurements discussed in this letter.

\section*{Acknowledgements}
We thank Takahiro Sagawa for useful discussions. 
%K.G.~is supported by JSPS KAKENHI Grant-in-Aid for Early-Career Scientists (21K13930) and Research Fellowships of Japan Society for the Promotion of Science for Young Scientists (22J00663).
M.N.~is supported by funds from the University of Chinese Academy of Sciences (UCAS), funds from the Kavli
Institute for Theoretical Sciences (KITS).
S.N.~is supported by MEXT KAKENHI (Grant Nos. 20H01838 and 21H05185).
R.Y.~is partially supported by MEXT KAKENHI  (Grant Nos.~20H01838, and 25K07156) and the WPI program ``Sustainability with Knotted Chiral Meta Matter (SKCM$^2$)'' at Hiroshima University.

\bibliography{cited}
\bibliographystyle{apsrev4-1}

\appendix

%\section{Classical and continuum limit} \label{app:explanation_spin_chirality}

\section{System size dependence}
In this appendix, we present the analysis of entanglement entropy, mutual information, and logarithmic negativity for various system sizes in the context of the $\mathrm{CI}$ model and random pure states. For the entanglement entropy calculations, the subsystem size is set to half of the total system size. When the total system size is odd, the subsystem size is determined by using the floor function. 
The system is divided into regions A, B, and C, consistent with the main text.

\begin{figure}[h]
    \centering
    \includegraphics[width=0.8\columnwidth]{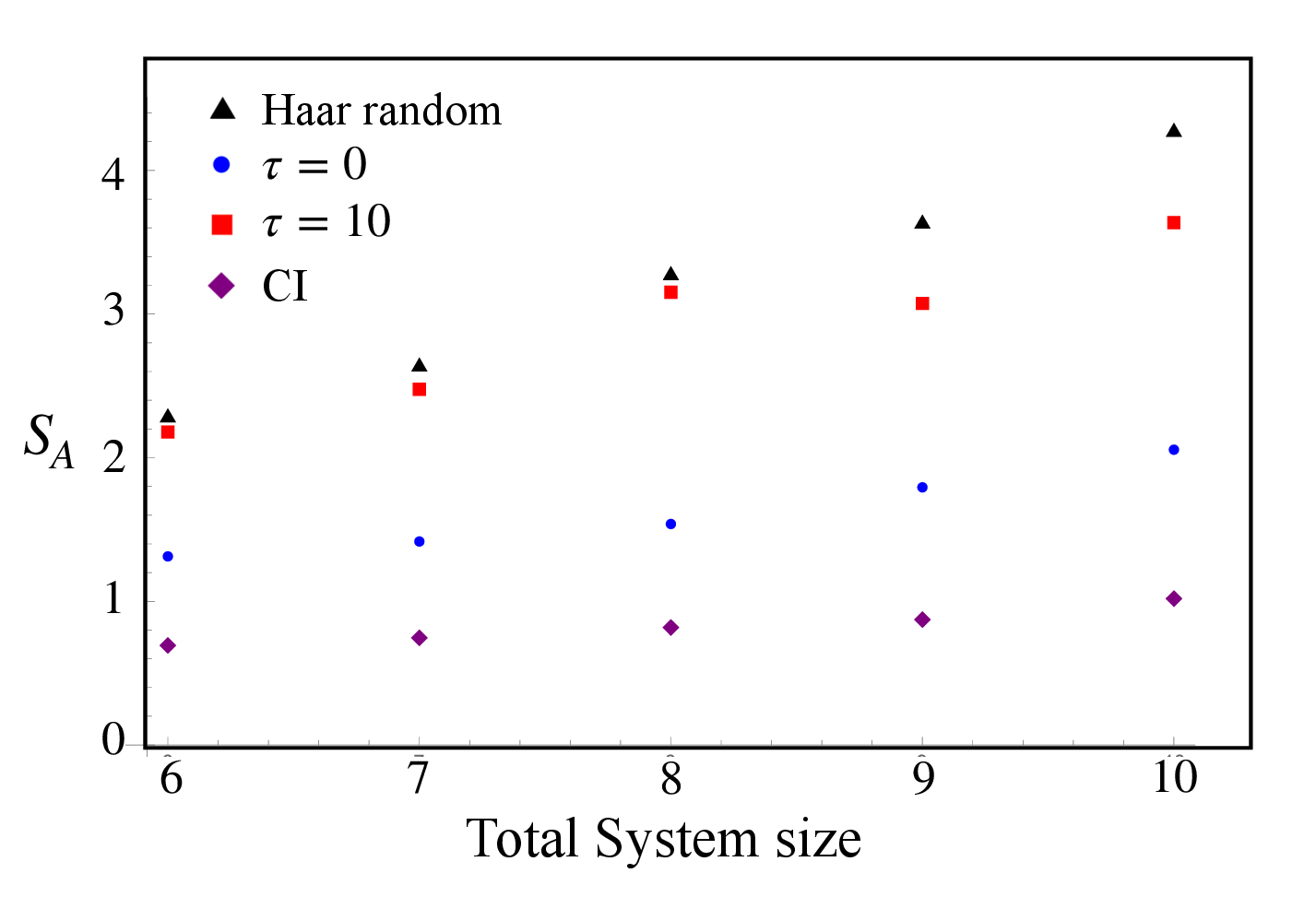}
    \caption{Time-averaged entanglement entropy as a function of total system size for $\tau = 0$, $\tau = 10$, and the $\mathrm{CI}$ model. The entanglement entropy for a random pure state is labeled as ``Haar random.''
    }
    \label{EEsizedep}
\end{figure}

In Fig.~\ref{EEsizedep}, we plot the time-averaged entanglement entropy as a function of total system size for $\tau = 0$, $\tau = 10$, and the $\mathrm{CI}$ model. The $\mathrm{CI}$ model exhibits a milder slope with respect to system size compared to the random pure state, labeled as ``Haar random'' in the figures.

\begin{figure}[h]
    \centering
    \includegraphics[width=0.8\columnwidth]{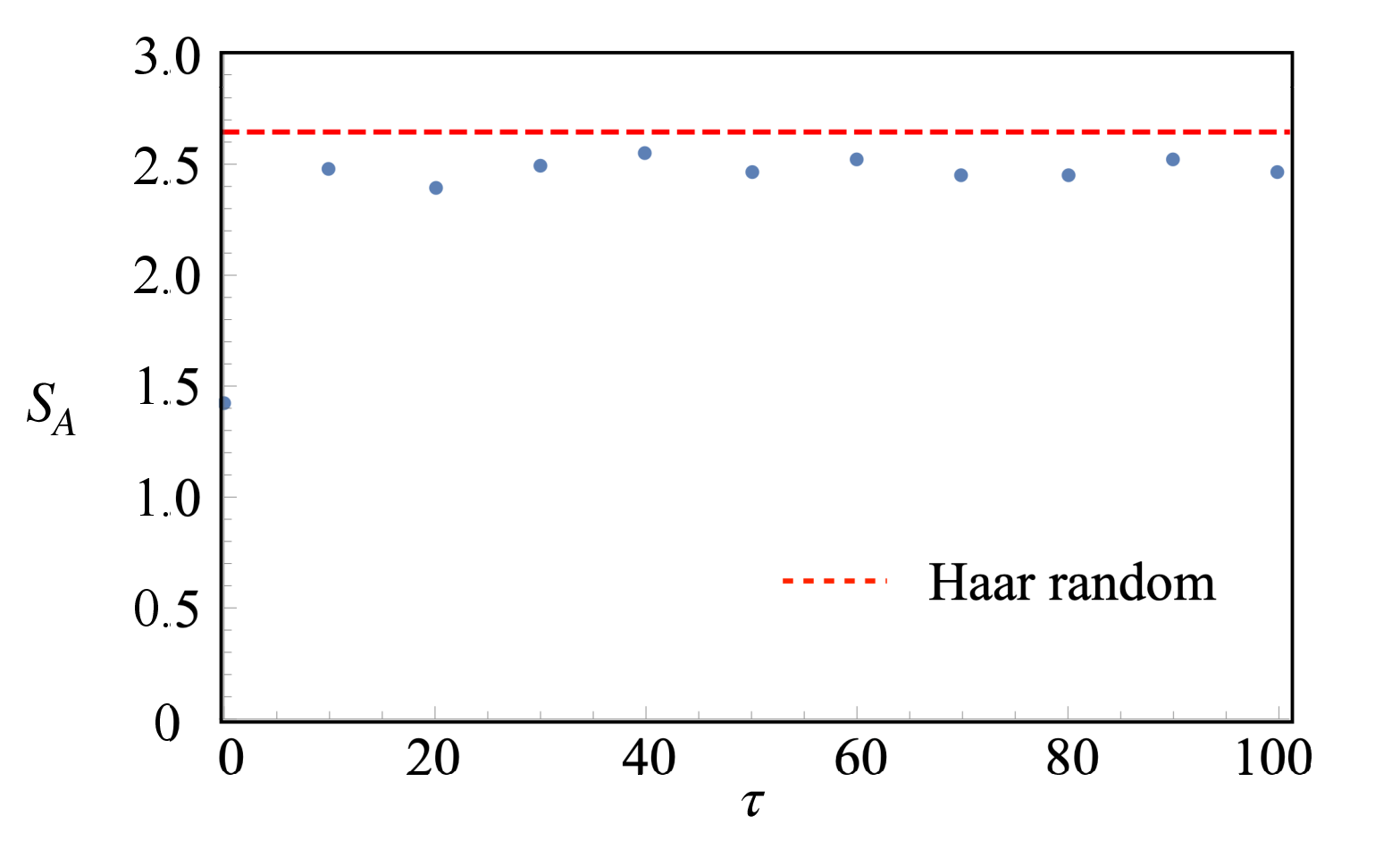}
    \caption{
    Long-time average value of the entanglement entropy as a function of $\tau$ for the $\mathrm{CI}$ model, compared to a random pure state (dashed line). 
    }
    \label{EEtaudep}
\end{figure}

In Fig.~\ref{EEtaudep}, we show the long-time average value of the entanglement entropy as a function of $\tau$. We set the system size to 7. The entanglement entropy approaches values close to those obtained for a random pure state, indicated by a dashed line. The non-monotonic behavior can be attributed to oscillatory dynamics in the system.

\begin{figure}[h]
    \centering
    \includegraphics[width=0.8\columnwidth]{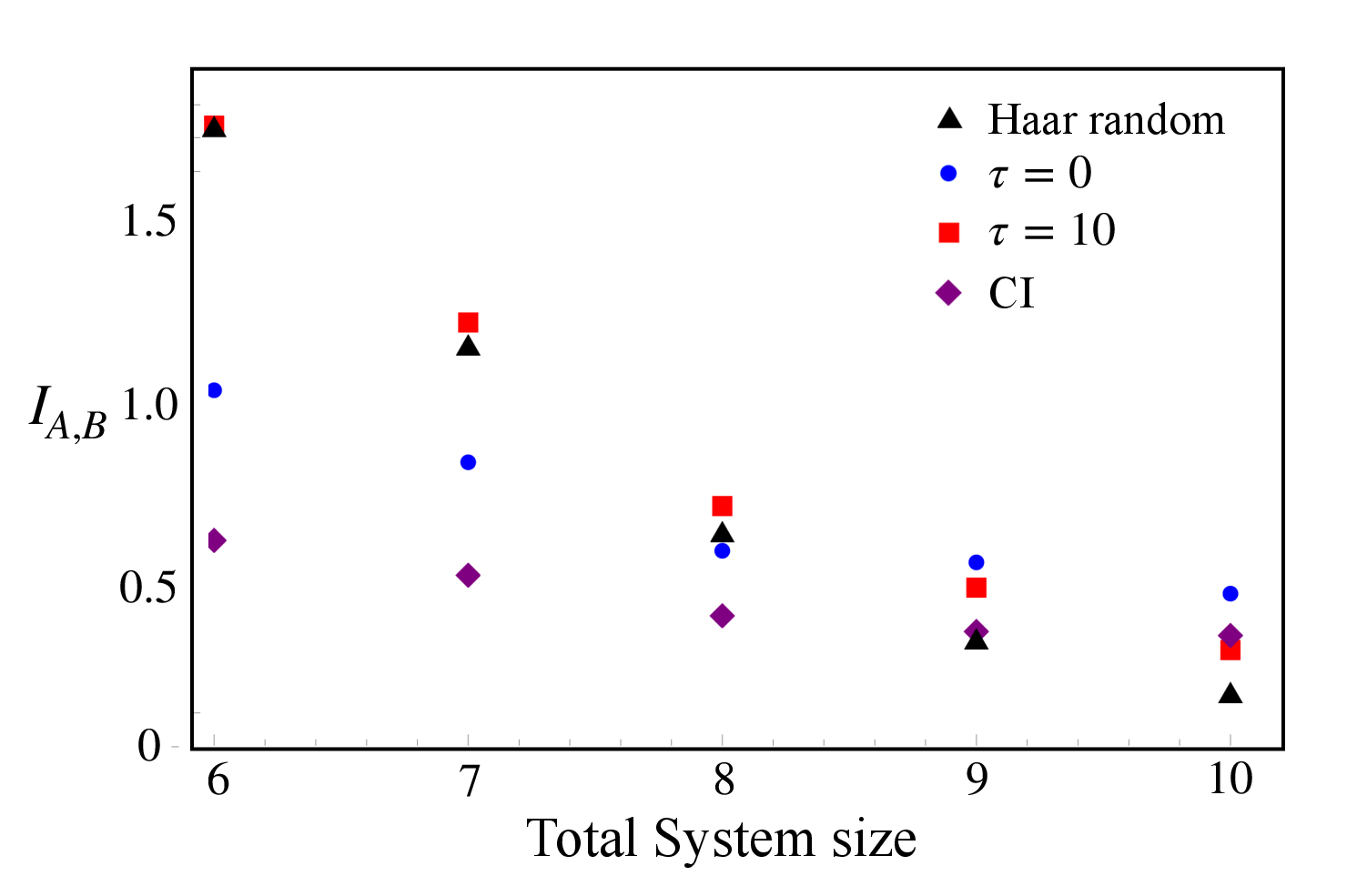}
    \caption{
    Long-time average value of the mutual information as a function of total system size for $\tau = 0$, $\tau = 10$, and the $\mathrm{CI}$ model. The entanglement entropy for a random pure state is labeled as ``Haar random.'' 
    }
    \label{MIsizedep}
\end{figure}

We also calculated the total size dependence of the long-time averaged mutual information. 
In Fig.~\ref{MIsizedep}, we present the long-time average value of the mutual information as a function of total system size for $\tau = 0$, $\tau = 10$, and the $\mathrm{CI}$ model. Similar to the entanglement entropy, the $\mathrm{CI}$ model shows a milder slope with respect to system size compared to the random pure state, labeled as ``Haar random.''

\begin{figure}[h]
    \centering
    \includegraphics[width=0.8\columnwidth]{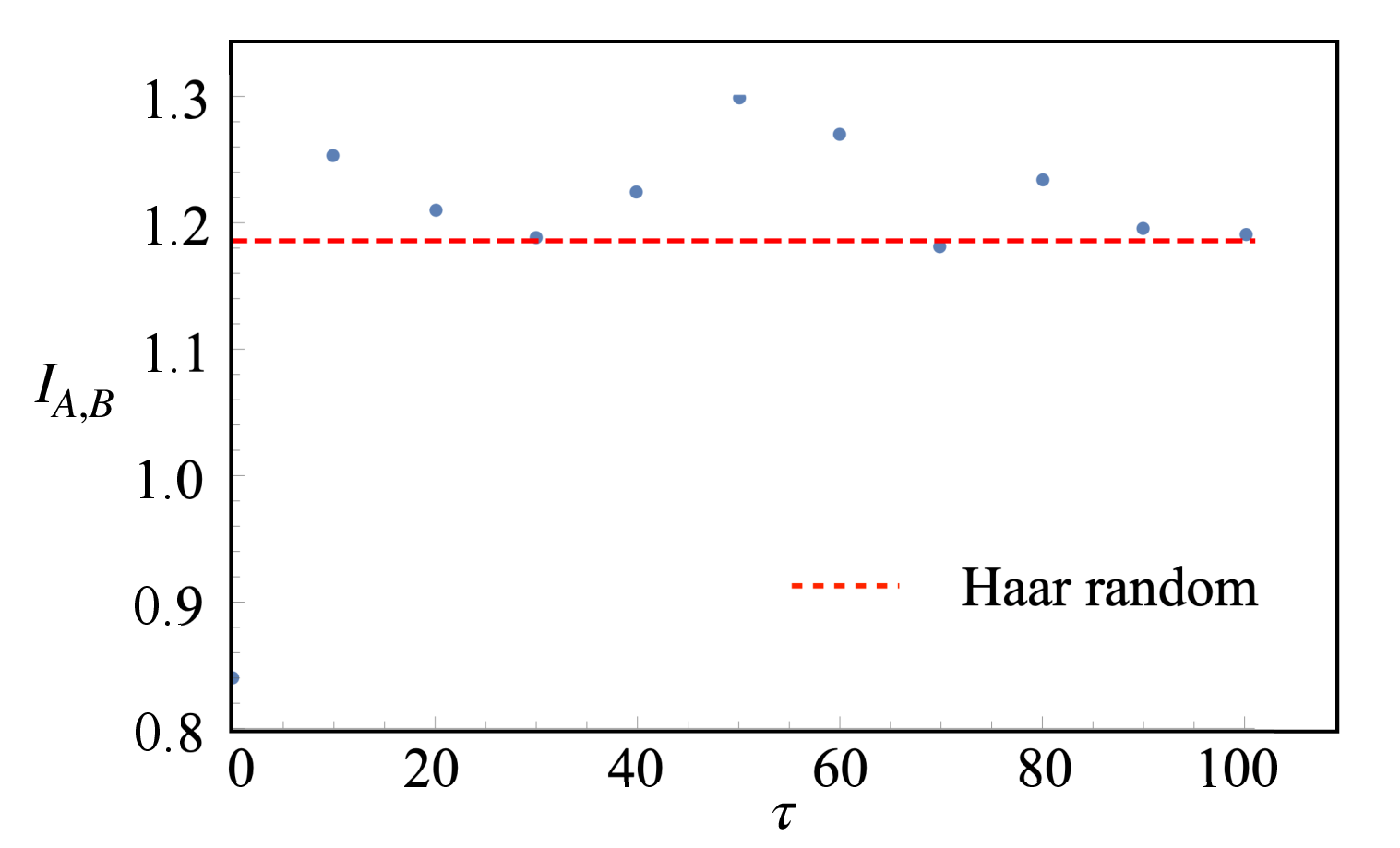}
    \caption{
    Long-time average value of the mutual information as a function $\tau$. The system size is $7$ and the partition of the system and the subsystem is the same with the one in the main text. 
    }
    \label{MItaudep}
\end{figure}

In Fig.~\ref{MItaudep}, we show the $\tau$ dependence of the long-term average value of the mutual information. 
Similar to the result for the entanglement entropy, the mutual information oscillates at the vicinity of the value obtained by the random pure state.

\begin{figure}[h]
    \centering
    \includegraphics[width=0.8\columnwidth]{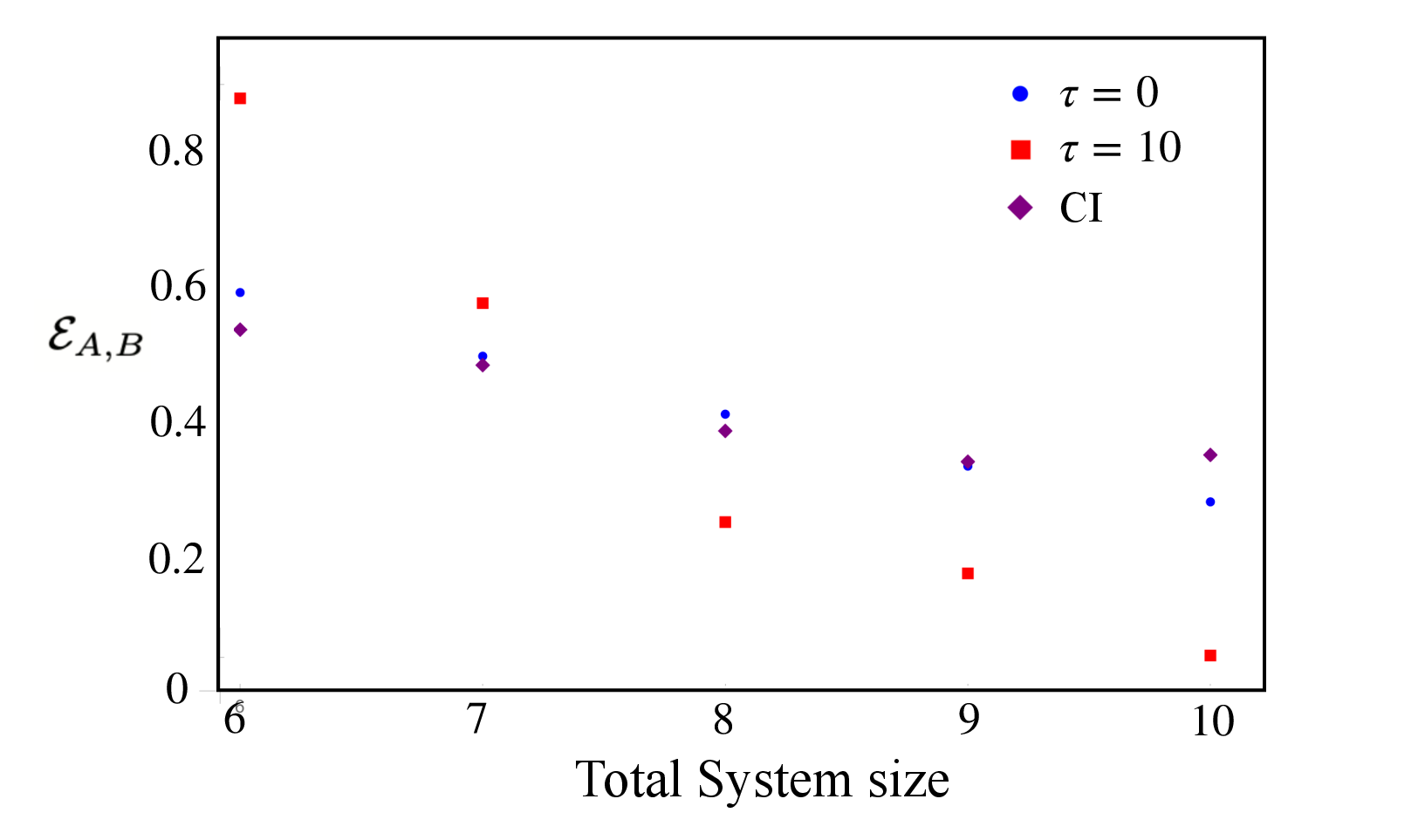}
    \caption{Time-averaged logarithmic negativity as a function of total system size for $\tau = 0$, $\tau = 10$, and the $\mathrm{CI}$ model. The entanglement entropy for a random pure state is labeled as ``Haar random.'' 
    }
    \label{LNsizedep}
\end{figure}

In Fig.~\ref{LNsizedep}, we illustrate the time-averaged logarithmic negativity as a function of total system size for $\tau = 0$, $\tau = 10$, and the $\mathrm{CI}$ model. The results are compared to those of a random pure state, labeled as ``Haar random.'' The system size and subsystem partitioning are consistent with those described in the main text.

\section{Long-time average} 
Due to the finite system size, temporal fluctuations in entanglement measures are observed. To facilitate comparison with the Haar random ensemble, we compute the long-time average of the entanglement entropy, defined as:
\begin{equation}
\bar{S}_A = \frac{1}{T - t_0} \int_{t_0}^T dt \, S_A(t),
\end{equation}
where $t_0 \ll T$, and $t_0$ is chosen to be larger than the initial relaxation time to exclude transient effects. In numerical calculations, this integral is approximated by a discrete summation over time steps. 
For mutual information and logarithmic negativity, the same procedure is used to obtain the long-time averaged value. 
For actual calculations, we set $t_0=10$ and $T=1000$.

\end{document}